\documentclass{iau} 
\usepackage{graphicx}
\usepackage{natbib}
\usepackage{overpic}

\newcommand{\LEAVEOUT}[1]{}

\title[] 
{Forecasting Solar Energetic Particle Fluence with Multi-Spacecraft Observations}

\author[Laitinen et al]   
{T. Laitinen$^{1}$, S. Dalla$^{1}$, M. Battarbee$^{1}$ and M.~S. Marsh$^{2}$}

\affiliation{
$^1$Jeremiah Horrocks Institute, University of Central Lancashire, Preston, United Kingdom.\\ email: {\tt tlmlaitinen@uclan.ac.uk} \\[\affilskip]
$^2$Met Office, Exeter, UK}

\pubyear{2017}
\volume{335}  
\jname{Space Weather of the Heliosphere: Processes and Forecasts}
\editors{C. Foullon \& O. Malandraki, eds.}
\begin{document}

\maketitle

\begin{abstract}
Forecasting Solar Energetic Particle (SEP) fluence, as integrated over an SEP event, is an important element when estimating the effect of solar eruptions on humans and technology in space. Current real-time estimates are based on SEP measurements at a single location in space. However, the interplanetary magnetic field corotates with the Sun approximately 13$^\circ$ each day with respect to Earth, thus in 4 days a near-Earth spacecraft will have changed their connection about 60$^\circ$ from the original SEP source. We estimate the effect of the corotation on particle fluence using a simple particle transport model, and show that ignoring corotation can cause up to an order of magnitude error in fluence estimations, depending on the interplanetary particle transport conditions. We compare the model predictions with STEREO observations of SEP events.
\end{abstract}

\firstsection 
\section{Introduction}

Total dose due to Solar Energetic Particles (SEPs) is an important factor in evaluation of radiation risk for astronauts and spacecraft. SEP fluence is affected by acceleration at the Sun and in interplanetary space \citep[e.g.][]{Reames1999}, scattering of particles along and across the field lines \citep[e.g.][]{Jokipii1966}, cross-field propagation due to field-line meandering \citep{LaEa2016parkermeand} and drifts \citep{Dalla2013}, and motion of the observer relative to the SEP source region, i.e. corotation \citep{Giajok2012,Marsh2015}. We study the effects of corotation on SEPs using a simple model and STEREO/LET (Mewaldt et al. 2008) SEP observations in 2013, when the spacecraft were at $\sim 60^\circ$ heliolongitudinal distance from each other. Such separation mimics the recently suggested L1/L5 mission concepts (e.g. Trichas et al. 2015).

\section{Effect of corotation of the SEP source: simple model}

\begin{figure}
  \begin{center}
    \includegraphics[width=0.9\linewidth]{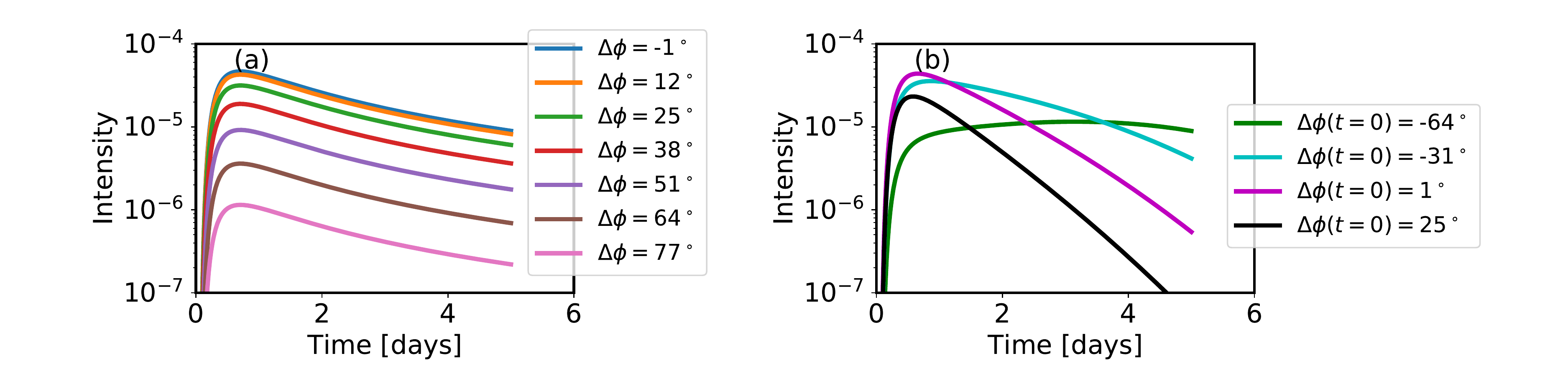}
\caption{ Time-intensity curves for 10 MeV protons given by Eq.~(\ref{eq:1ddiff}) with $\lambda=0.03$~AU at $r=$1~AU. (a): SEP intensities on field lines that corotate with the Sun. (b): SEP intensities observed with spacecraft at 1~AU moving relative to the corotating field lines.
\label{fig:corotfluxes}}
\end{center}
\end{figure}

We estimate the effect of the corotation of the particle source on the SEP intensities  with a simple 1D diffusion model. We use impulsive injection with Gaussian longitudinal injection distribution, which gives the particle intensity $n$ at the observing spacecraft as 
  \begin{equation}
    \label{eq:1ddiff}
    n(t,r,\phi)=n_0(\phi)\frac{1}{2\pi\left(\lambda v t/3\right)^{3/2}}
      \mathrm{e}^{-\frac{3r^2}{4\lambda v t}},\,\,\,\;\;\;\;\;
n_0(\phi)=n_{00}\,\mathrm{e}^{-(\Delta\phi)^2/(2\sigma_\phi^2)}
  \end{equation}
  where $\phi$ and $r$ are the observer's heliographic longitude and radial distance from the Sun, $t$ is time from injection, $\lambda$ the scattering mean free path, $v$ the SEP speed, $\Delta\phi=\phi_F-\phi_{S/C,F}$ with $\phi_F$ and $\phi_{S/C,F}$ the SEP source and spacecraft footpoint longitudes, $\sigma_\phi=40^\circ$ \citep[e.g.][]{Richardson2014}, and $n_{00}$ a constant. We show in Figure~\ref{fig:corotfluxes}~(a) the SEP intensities at 1~AU for different constant $\Delta\phi$ values. 

A spacecraft observing the SEPs at 1~AU rotates at rate $\Omega=-13.2^\circ$/day with respect to the SEP surface, giving
$\Delta\phi=\phi_F-(\phi_{S/C,F}(t=0)+\Omega t)$. The spacecraft thus moves with respect to field lines that are connected to different-strength SEP injections at the Sun. The effect of this motion on SEP fluxes is demonstrated in Figure~\ref{fig:corotfluxes}~(b), where the initially well-connected spacecraft (magenta curve) sees rapid SEP intensity onset followed by fast decay, whereas a spacecraft seeing the SEP event as central (green curve) sees the SEP intensities remaining high for several days.  Similar features can be seen in SEP event profiles observed with different initial longitudinal separations (Cane et al. 1988). As seen in our simple model, corotation can produce these features without requiring a continuous, moving SEP source, as shown also by Marsh et al. (2015).

\section{Multi-spacecraft observations}

We analysed 10~MeV proton intensities during 10 events where SEPs were observed by both STEREO/A (STA) and STEREO/B (STB) LET instruments \citep{STEREOLET} in 2013. The source longitude $\phi_F$ was analysed from STEREO and SOHO observations, and compared with previous studies \citep[e.g.][]{Richardson2014}, while $\phi_{S/C,F}$ was obtained using the solar wind velocity at STA and STB.

\begin{figure}
  \begin{center}
    \includegraphics[width=0.45\linewidth]{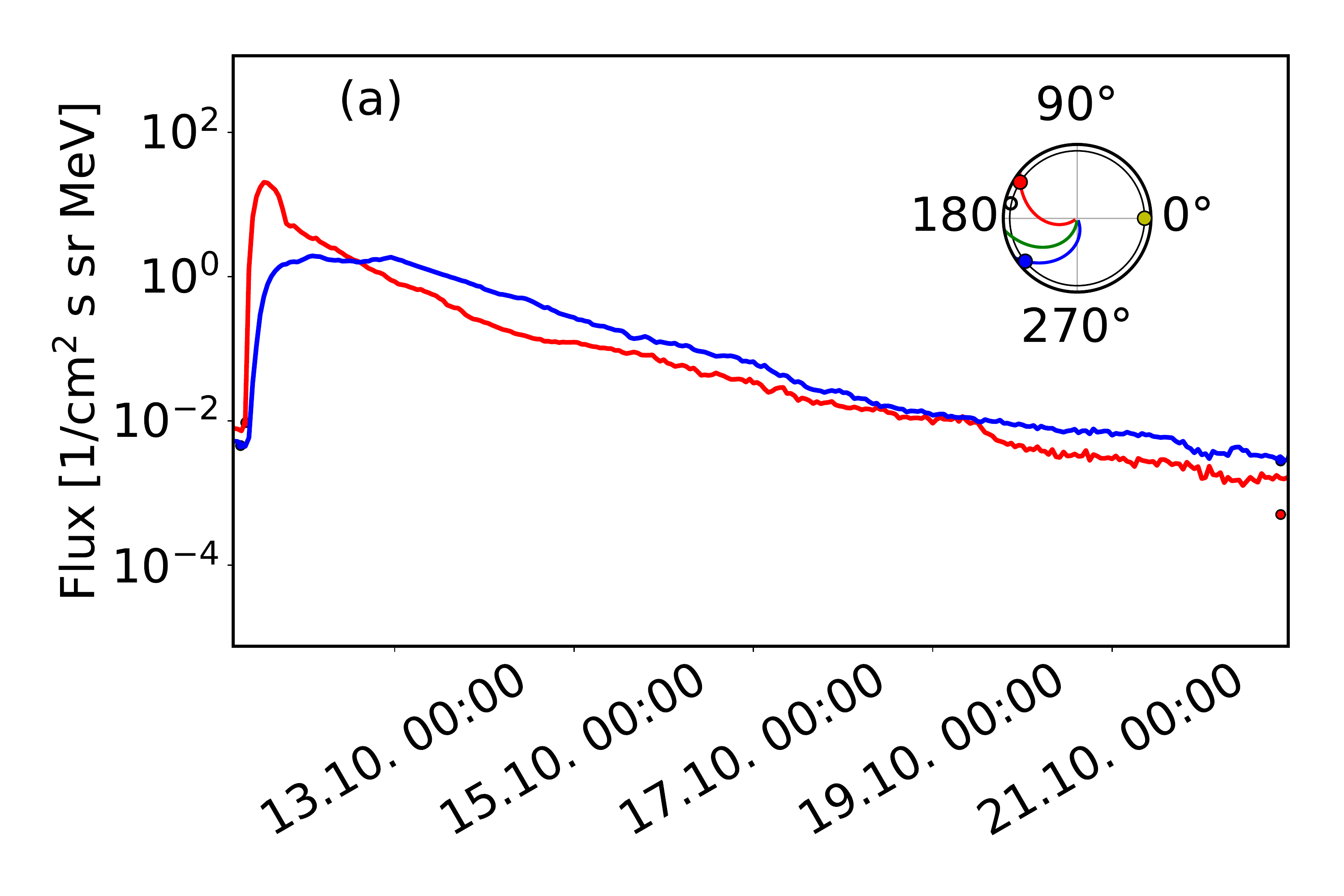}
    \includegraphics[width=0.45\linewidth]{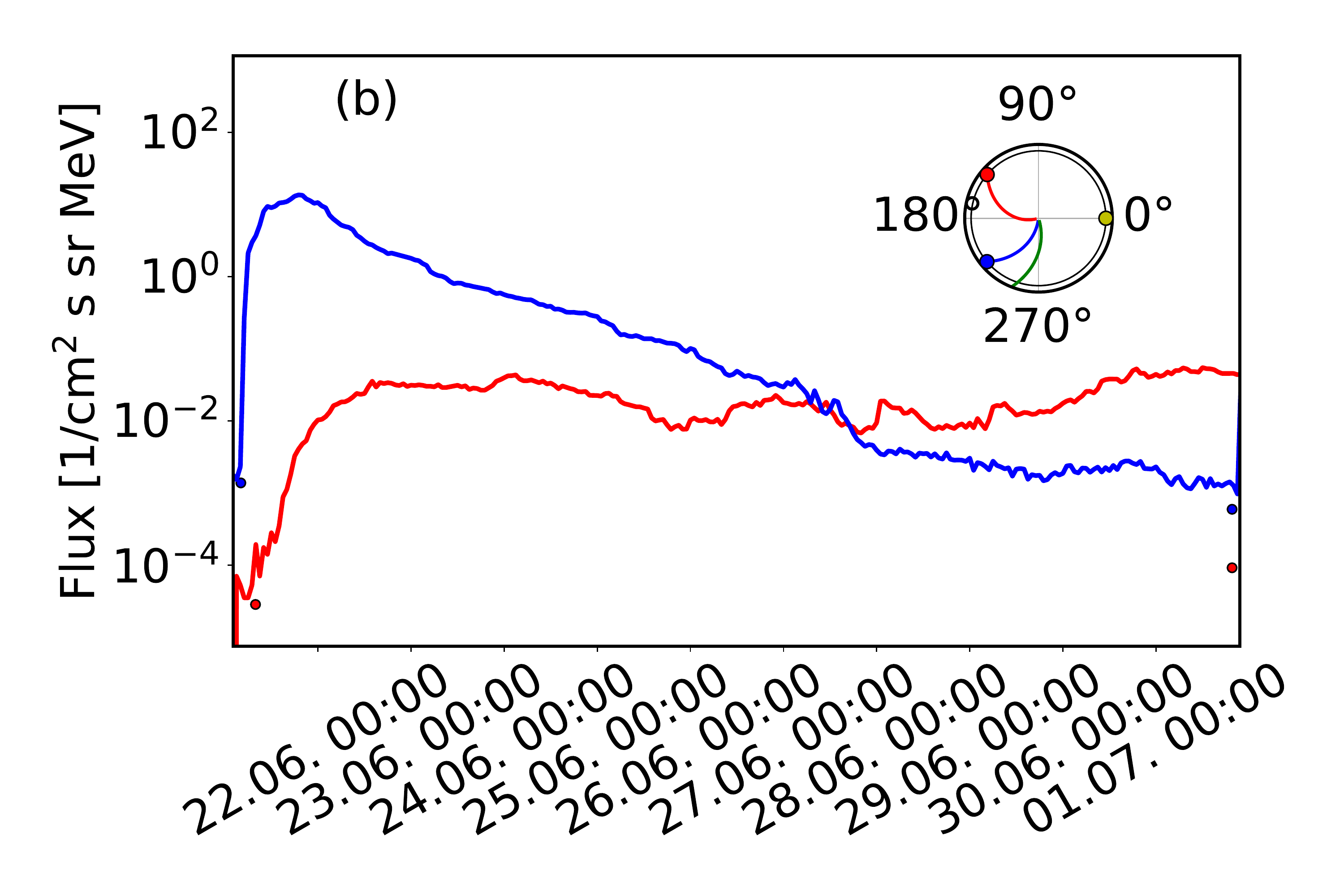}
\caption{ 10~MeV proton intensities on (a) 11 October and (b) 21 June 2013. In the inset, the red and blue circles show the STA and STB locations at the start of the event, respectively, and the red, blue
  and green spirals show the Parker spirals connected to the STA, STB
  and the SEP source location.
  \label{fig:STASTBevs}}
\end{center}
\end{figure}

Figure~\ref{fig:STASTBevs}~(a) shows an event with the SEP source between the two spacecraft. As with all (4/10) events with this connectivity, the slower increase at STB is qualitatively consistent with STB propagating towards higher-intensity field lines. Figure~\ref{fig:STASTBevs}~(b) shows an event where both spacecraft are advancing away from the source region. STA reaches $\Delta\phi=180^\circ$ on 27 June, which is seen as an increase in intensities. Overall, the SEP event profiles showed signatures of SEP source corotation.

\begin{figure}
  \begin{center}
    \includegraphics[width=0.38\linewidth]{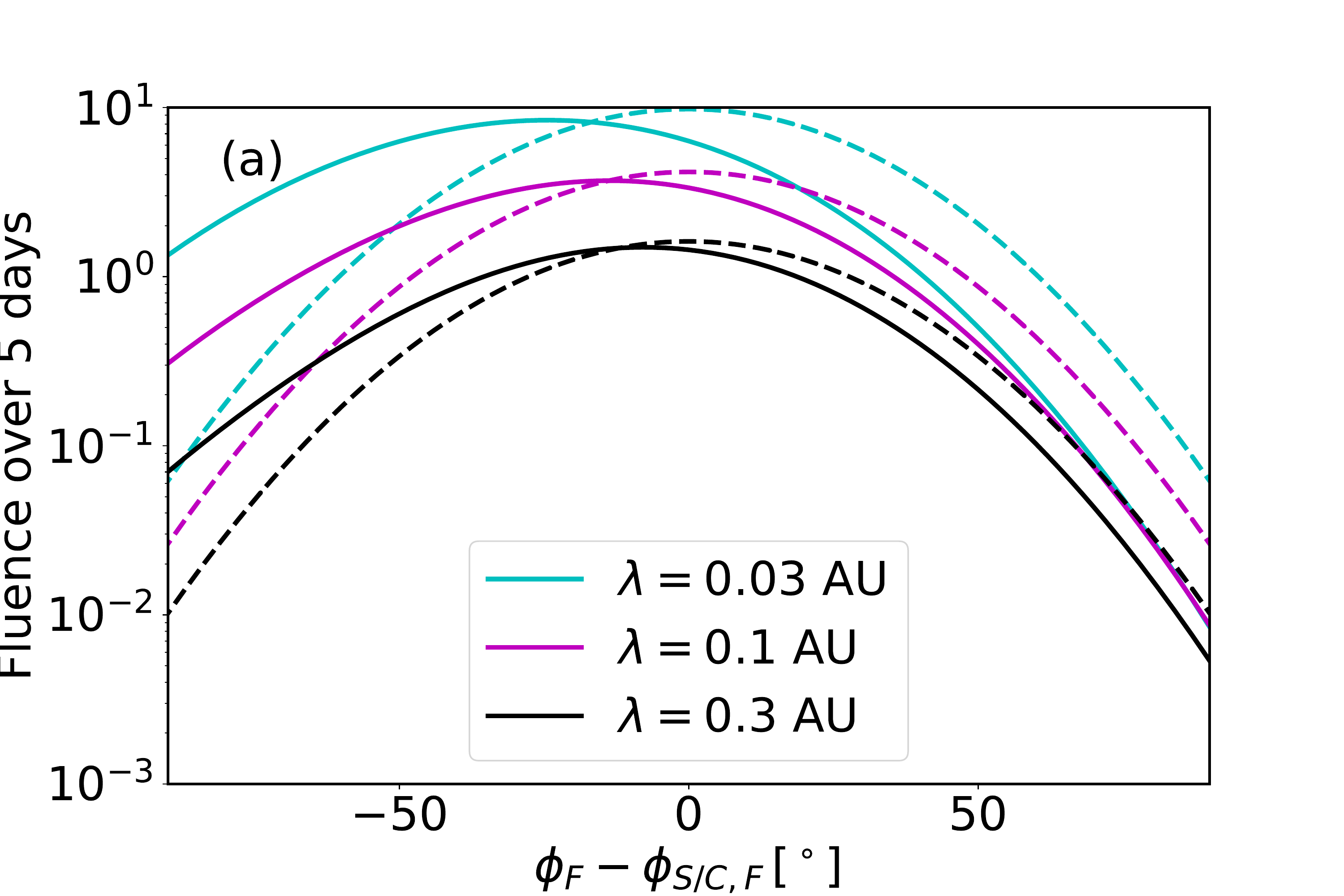}
    \includegraphics[trim=0cm 1cm 0cm 0cm,
    clip=true,width=0.38\linewidth]{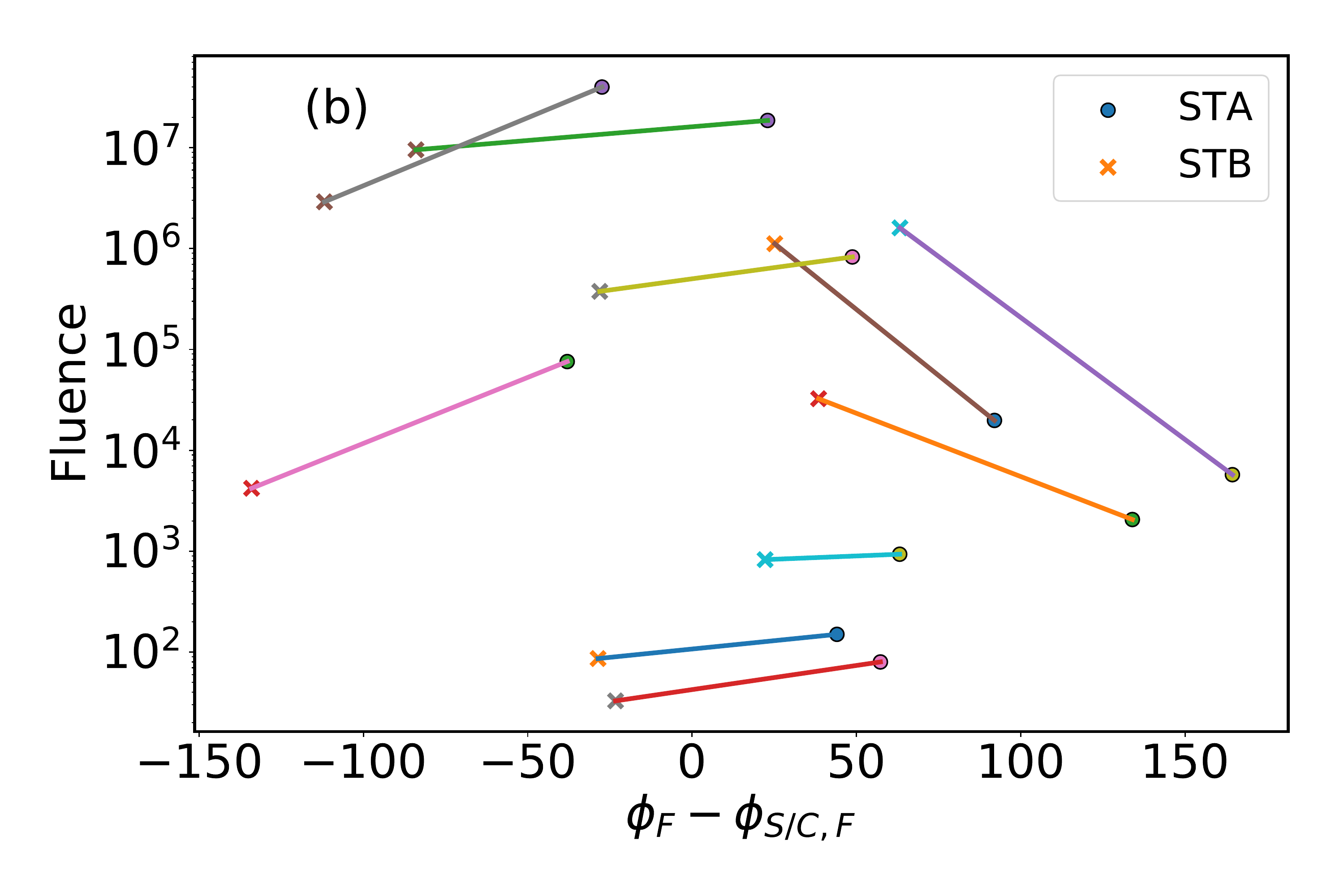}
\caption{\textbf{(a)} Modelled 10 MeV proton fluence at 1~AU from the Sun as a function of $\Delta\phi(t=0)$. The dashed curves show the fluence if
  solar rotation is ignored. \textbf{(b)}  Event-integrated fluences for 10 MeV 10 SEP events as a function of $\Delta\phi(t=0)$. %
\label{fig:fluences}}
\end{center}
\end{figure}

In Figure~\ref{fig:fluences}~(a), we show 5-day fluences of 10~MeV protons, modelled with Eq.~(\ref{eq:1ddiff}), as a function of the initial $\Delta\phi$, both for cases including corotation (solid curves) and ignoring corotation (dashed curves). The corotation causes a significantly increased fluence for initially eastern SEP sources ($\Delta\phi(t=0)<0$). The fluence also depends strongly on~$\lambda$.  In Figure~\ref{fig:fluences}~(b), we show the observed SEP fluences for the analysed events, as a function of the longitude. As can be seen, within a given event the fluences are generally larger for $\Delta\phi(t=0)\sim 0$, however there is considerable scatter, and low statistics prevent deducing clear asymmetries.

\section{Conclusions}

Propagation models show that the corotation of the SEP-filled field lines with the Sun affects the SEP intensities observed by interplanetary spacecraft. Our SEP event analysis shows that the time-intensity morphology is consistent with corotation affecting the event profiles. Low statistics prevent finding signatures of corotation in fluences without further, more involved modelling. 
Two-spacecraft observations bring more information, but require further study before their value can be estimated fully, including taking into account the varying interplanetary turbulence \citep[e.g.][]{Burlaga1976}, which has significant influence on SEP transport along and across the mean field, and SEP fluences.

\textbf{Acknowledgements.} We acknowledge the STEREO/IMPACT, STEREO/SECCHI and SOHO/LASCO teams for providing the data used in this study.  TL and SD acknowledge support from the UK Science and Technology Facilities Council (STFC) (grants ST/J001341/1 and ST/M00760X/1) and MB from the Leverhulme Trust Grant RPG-2015-094.



\begin{thebibliography}{11} \expandafter\ifx\csname natexlab\endcsname\relax\def\natexlab#1{#1}\fi \bibitem[{{Burlaga} \& {Turner}(1976)}]{Burlaga1976} {Burlaga}, L.~F. \& {Turner}, J.~M. 1976, JGR, 81, 73 \bibitem[{{Cane} {et~al.}(1988){Cane}, {Reames}, \& {von   Rosenvinge}}]{Cane1988} {Cane}, H.~V., {Reames}, D.~V., \& {von Rosenvinge}, T.~T. 1988, JGR, 93, 9555 \bibitem[{{Dalla} {et~al.}(2013){Dalla}, {Marsh}, {Kelly}, \&   {Laitinen}}]{Dalla2013} {Dalla}, S., {Marsh}, M.~S., {Kelly}, J., \& {Laitinen}, T. 2013, J. Geophys.   Res. (Space Physics), 118, 5979 \bibitem[{{Giacalone} \& {Jokipii}(2012)}]{Giajok2012} {Giacalone}, J. \& {Jokipii}, J.~R. 2012, ApJL, 751, L33 \bibitem[{{Jokipii}(1966)}]{Jokipii1966} {Jokipii}, J.~R. 1966, ApJ, 146, 480 \bibitem[{{Laitinen} {et~al.}(2016){Laitinen}, {Kopp}, {Effenberger}, {Dalla},   \& {Marsh}}]{LaEa2016parkermeand} {Laitinen}, T., {Kopp}, A., {Effenberger}, F., {Dalla}, S., \& {Marsh}, M.~S.   2016, A\&A, 591 \bibitem[{{Marsh} {et~al.}(2015){Marsh}, {Dalla}, {Dierckxsens}, {Laitinen}, \&   {Crosby}}]{Marsh2015} {Marsh}, M.~S., {Dalla}, S., {Dierckxsens}, M., {Laitinen}, T., \& {Crosby},   N.~B. 2015, Space Weather, 13, 386 \bibitem[{{Mewaldt} {et~al.}(2008){Mewaldt}, {Cohen}, {Cook}, {Cummings},   {Davis}, {Geier}, {Kecman}, {Klemic}, {Labrador}, {Leske}, {Miyasaka},   {Nguyen}, {Ogliore}, {Stone}, {Radocinski}, {Wiedenbeck}, {Hawk}, {Shuman},   {von Rosenvinge}, \& {Wortman}}]{STEREOLET} {Mewaldt}, R.~A., {Cohen}, C.~M.~S., {Cook}, W.~R., {et~al.} 2008, Space Sci.   Rev., 136, 285 \bibitem[{{Reames}(1999)}]{Reames1999} {Reames}, D.~V. 1999, Space Sci. Rev., 90, 413 \bibitem[{{Richardson} {et~al.}(2014){Richardson}, {von Rosenvinge}, {Cane},   {Christian}, {Cohen}, {Labrador}, {Leske}, {Mewaldt}, {Wiedenbeck}, \&   {Stone}}]{Richardson2014} {Richardson}, I.~G., {von Rosenvinge}, T.~T., {Cane}, H.~V., {et~al.} 2014,   Sol. Phys., 289, 3059 \bibitem[{{Trichas} {et~al.}(2015){Trichas}, {Gibbs}, {Harrison}, {Green},   {Eastwood}, {Bentley}, {Bisi}, {Bogdanova}, {Davies}, {D'Arrigo}, {Eyles},   {Fazakerley}, {Hapgood}, {Jackson}, {Kataria}, {Monchieri}, \&   {Windred}}]{CarringtonMission2015} {Trichas}, M., {Gibbs}, M., {Harrison}, R., {et~al.} 2015, Hipparchos, vol.~2,   Issue 12, pp.~25 - 31, 2, 25 \end{thebibliography}

\end{document}